\documentclass[aps, pra, reprint, letterpaper, longbibliography, superscriptaddress]{revtex4-1}

\usepackage{amsmath,amssymb,amsfonts,latexsym,color,dcolumn,bm,mathtools,amsbsy}
\usepackage[english]{babel}
\usepackage{graphicx}
\usepackage[utf8]{inputenc}
\usepackage{textcomp}
\usepackage{siunitx}
\usepackage[colorlinks,urlcolor=blue,citecolor=blue,unicode]{hyperref}
\usepackage{verbatim}

\newcommand*{\red}{\textcolor{red}}

\begin{document}

\title{Squeeze film pressure sensors based on SiN membrane sandwiches}

\author{Sepideh Naserbakht}
\author{Aur\'{e}lien Dantan}
\email{dantan@phys.au.dk}
\affiliation{Department of Physics and Astronomy, University of Aarhus, DK-8000 Aarhus C, Denmark}

\begin{abstract}
We realize squeeze film pressure sensors using suspended, high mechanical quality silicon nitride membranes forming few-micron gap sandwiches. The effects of air pressure on the mechanical vibrations of the membranes are investigated in the range $10^{-3}-50$ mbar and the intermembrane coupling induced by the gas is discussed in light of a squeeze film coupled-oscillator model. The high responsivity (several kHz/mbar) and the sub-pascal sensitivity of such simple pressure sensors are attractive for absolute and direct pressure measurements in rarefied air or high vacuum environments.
\end{abstract}

\date{\today}

\maketitle


\section{Introduction}

Conventional miniaturized pressure sensors, such as capacitive or piezo-resistive sensors, are based on either a static determination of pressure by measuring the deflection of a suspended membrane due to a pressure difference with a reference cavity or dynamical  changes in resonance frequencies~\cite{Bhat2007} and typically operate from atmospheric pressure to low vacuum (milllibar). For high and ultrahigh vacuum measurements, ionization gauge sensors commonly exhibit a high pressure sensitivity, but require knowledge of the involved species and their ionization cross sections. In contrast, squeeze film pressure sensors~\cite{Bao2000,Ekinci2005,Ekinci2010}--based on the dynamical modification of the mechanical properties of oscillating elements due to the compression of a fluid in a small gap region--allow in principle absolute and direct pressure measurements. In the rarefied gas (or free molecular flow) regime, in which the mean free path of the gas molecules exceeds the gap dimension, intermolecular collisions are rare and oscillator-molecule collisions predominant; while the collisions of the molecules contribute to dampen the mechanical oscillations, the spring constant added by the squeezed gas can substantially modify the mechanical resonance frequencies of the oscillator~\cite{Blech1983,Veijola1995,Bao2000,Bao2007}. Interestingly, for high enough oscillation frequencies, the gas-added spring constant is directly proportional to the pressure and may enable species-independent pressure measurements. Squeeze film effects have thus been investigated with various oscillators and regimes~\cite{Prak1991,Blom1992,Andrews1993,Legtenberg1994,Steeneken2004,Vignola2006,Verbridge2008,Mol2009,Suijlen2009,Southworth2009,Stifter2012,Suijlen2012,Kainz2014,Kumar2015,Dolleman2016,Naesby2017}.

Suspended dielectric plates with small thickness/large area and possessing high quality factor mechanical resonances are well-suited oscillators for investigating the effects of pressure--whether due radiation or the interaction with a fluid. For the latter, few tens of nanometers-thick, tensioned silicon nitride (SiN) membranes constitute excellent and versatile mechanical resonators, as their low loss level allows for efficiently coupling them to electromagnetic fields~\cite{Thompson2008} and their outstanding mechanical properties enable the observation of a wide variety of optomechanical effects in the context of cavity optomechanics~\cite{Aspelmeyer2014}. Arrays of suspended membranes~\cite{Nair2017,Piergentili2018,Gartner2018,Wei2019} are particularly interesting in that respect, as they allow for enhancing radiation pressure forces~\cite{Xuereb2012,Xuereb2013} and open for mediating interactions between the resonators and exploiting collective optomechanical phenomena~\cite{Bhattacharya2008,Hartmann2008,Seok2012,Xuereb2014,Kipf2014,Xuereb2015,Bemani2017}.

Such properties also make them naturally attractive for investigating fluid pressure variations. The use of a membrane sandwich in which a fluid is compressed in the small gap region between the membranes makes it possible to exploit the squeeze film effect in order to determine the gas pressure and couple the membranes together~\cite{Naesby2017}. We report here on the realization of squeeze film pressure sensors based on suspended, high-mechanical quality SiN membranes forming a small gap sandwich and characterize the modifications of their mechanical properties in air in the range $10^{-3}- 50$ mbar. Pressure-induced frequency shifts as high as 4 kHz/mbar are demonstrated, a substantial improvement over previous arrays with larger gaps~\cite{Naesby2017}. Such a pressure responsivity is close to the highest (9 kHz/mbar) reported with high frequency graphene microdrums~\cite{Dolleman2016}, and substantially extends the pressure range and sensitivity of this type of sensors into the sub-pascal regime. Such performances result from a combination of several factors: the small gap (2-3 $\mu$m) and the membrane thickness ($<100$ nm) make for strong squeeze film effecs, the high tensile stress allows for operating with high-frequency ($\sim$ MHz) and high-Q ($\sim 10^5$) modes, the large area (0.25 mm$^2$) ensures trapping of the gas and an essentially elastic squeeze film force, and, last, the squeeze film-induced couplings in the chosen sandwich geometry allows for increasing the pressure responsivity. Such squeeze film pressure sensors would be relevant for a wide range of applications ranging from absolute high vacuum pressure calibration~\cite{Volklein2007,Gorecka2009} or the direct determination of the vapor pressure of chemically or environmentally relevant substances~\cite{Bilde2015}.

The paper is outlined as follows: Sec.~\ref{sec:squeeze} discusses the dynamics of high-frequency vibrating plates in the free molecular flow regime and introduces the theoretical coupled-mode model used to analyze these dynamics in the specific case of a membrane sandwich structure. Section~\ref{sec:sandwich} discusses the assembly and optical characterization of these arrays and describes the experimental setup used for the characterization of their vibrations. Section~\ref{sec:results} presents the results of measurements of the effects of air pressure on the mechanical properties of two membrane sandwiches having gaps between 2 and 3 microns. In Sec.~\ref{sec:conclusion} we conclude and discuss the prospects for improving the responsivity and sensitivity of these squeeze film sensors.

\section{Squeeze film effects in a membrane sandwich}
\label{sec:squeeze}

\begin{figure}[h]
\begin{flushleft}\includegraphics[width=0.9\columnwidth]{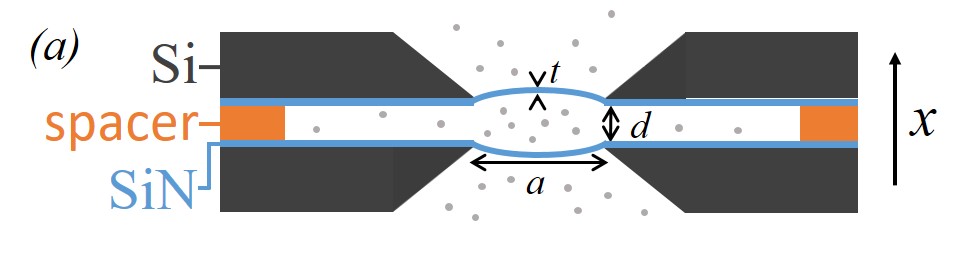}\end{flushleft}
\includegraphics[width=0.95\columnwidth]{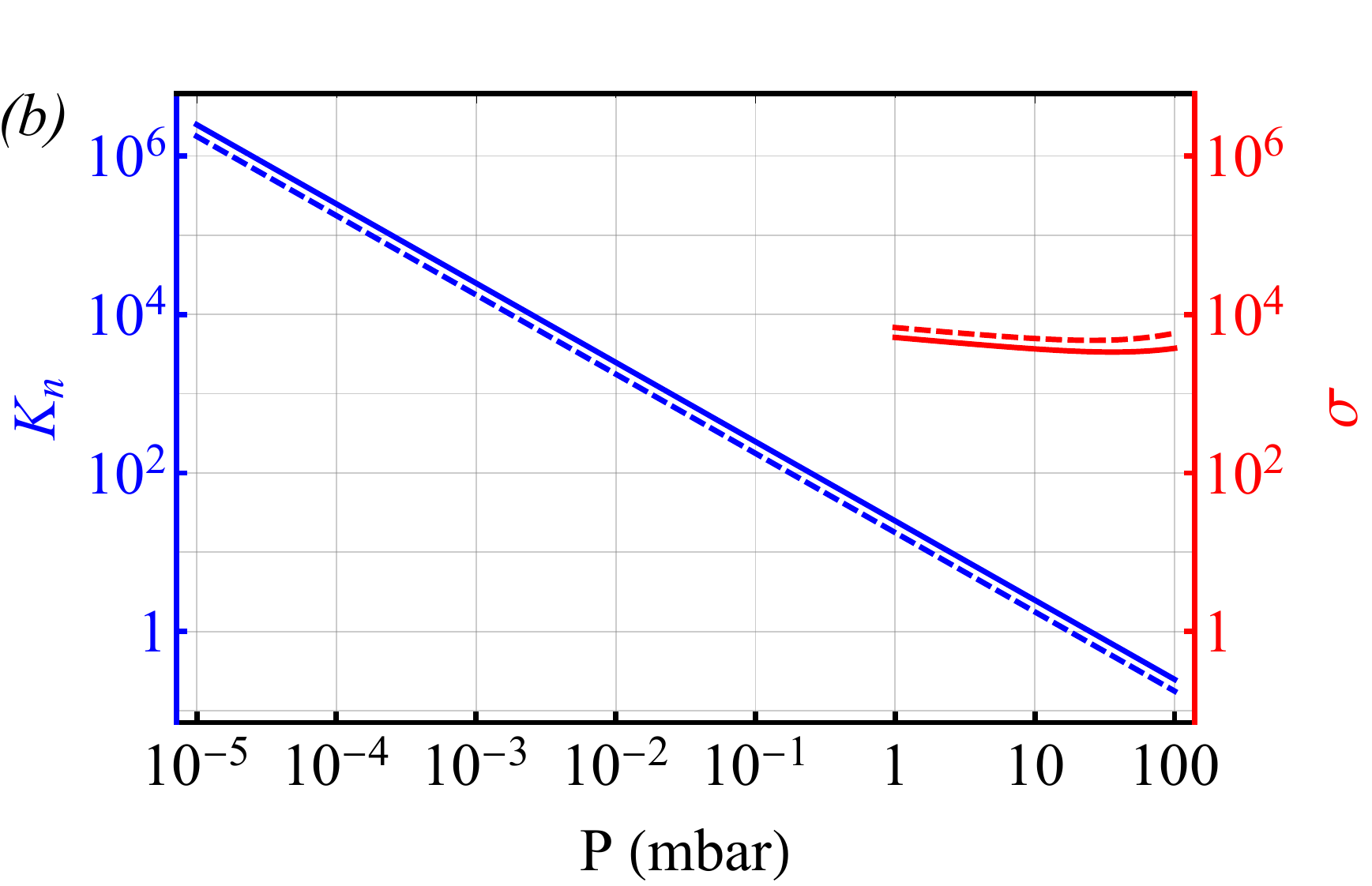}
\caption{(a) Cross section schematic of the membrane sandwich (not to scale). (b) Variations with pressure of the Knudsen number $K_n$ (blue) and squeeze parameter $\sigma$ (red), for sandwiches with an intermembrane separation $d$ of 2.1 $\mu$m (plain) and 2.95 $\mu$m (dashed).}
\label{fig:sandwich}
\end{figure}

We consider a membrane sandwich consisting in two parallel thin clamped plates with lateral dimension $a$, thickness $t$ and separated by a distance $d$ ($a\gg d,t$), as depicted in Fig.~\ref{fig:sandwich}. The membranes are surrounded by air at ambient pressure $P$ on all sides. We focus here on the rarefied air regime, which is characterized by a high Knudsen number,
\begin{equation}
K_n=\frac{\lambda}{d}=\frac{k_BT}{\sqrt{2}\pi \sigma_{\textrm{air}}^2Pd},
\end{equation}
defined as the ratio of the mean free path and the gap dimension, where $k_B$ is the Boltzmann constant, $T$ the temperature and $\sigma_{\textrm{air}}=4.19\times 10^{-10}$ m the air molecule diameter. The variations of $K_n$ in the pressure range investigated experimentally in the following are shown in Fig.~\ref{fig:sandwich}b, for sandwiches with intermembrane separations $d$ of 2.1 and 2.95 $\mu$m, respectively. The Knudsen number becomes of order unity for pressures of order ten or a few tens of millibars, indicating the transition to the viscous fluid regime. In the rarefied air regime, the collisions of the air molecules with the membranes give rise to an additional damping, proportional to the pressure~\cite{Christian1966} and given by
\begin{equation}
\gamma_{\textrm{air}}=4\sqrt{\frac{2}{\pi}}\sqrt{\frac{M_{\textrm{air}}}{RT}}\frac{P}{\rho t},
\label{eq:gamma_air}
\end{equation}
where $M_{\textrm{air}}=29$ g/mol is the air molar mass, $R=8.31$ J/K/mol the ideal gas constant and $\rho=2700$ kg/m$^3$ the density of SiN. The isothermal compression of the gas between the membranes gives an additional force on the membranes (squeeze film effect)~\cite{Bao2007}. For a fluid with a relaxation time $\tau$ interacting with a resonator with angular frequency $\omega$, the Newtonian hydrodynamics assumption that the fluid remains in the vicinity of equilibrium is valid when the Weissenberg number, $W_i=\omega\tau$, is much larger than unity~\cite{Ekinci2010}. The relaxation time corresponds to the average time it takes the molecules to leave the gap, which can be estimated by~\cite{Suijlen2009}
\begin{equation}
\tau=\frac{8a^2}{\pi^3 d\bar{v}},
\end{equation}
where $\bar{v}=\sqrt{8RT/\pi M_{\textrm{air}}}$ is the mean velocity of the air molecules. For $d=2.1$ $\mu$m, $T=298$ K and a fundamental mode frequency $\omega/(2\pi)=820$ kHz, one gets $W_i\simeq 340$; the squeeze film force is then expected to be predominantly elastic in the rarefied air regime. 

Since the transition regime is also relevant for the pressure range investigated here, one can consider the squeeze film dynamics in the opposite, viscous fluid regime. There, the relevant parameter to characterize the fluid behavior is the squeeze number parameter~\cite{Blech1983},
\begin{equation}
\sigma=\frac{\pi d^2 P}{24a^2\mu},
\end{equation}
where $\mu$ is the air viscosity. In order to get an estimate of $\sigma$ , we use the empirical effective viscosity of~\cite{Veijola1995},
$\mu_{\textrm{eff}}=\mu_0/(1+9.658 K_n^{1.159})$,
where $\mu_0=1.8\times 10^{-5}$ Pa is the air viscosity at atmospheric pressure, and get that $\sigma\simeq 3600$ for $d=2.1$ $\mu$m and $P=10$ mbar (see also Fig.~\ref{fig:sandwich}b). One can thus reasonably expect the squeeze film force to be essentially elastic for the whole pressure range considered in this work. 

In this regime and for a large oscillating plate closely lying with a fixed plate, the air-added spring constant is proportional to the pressure and results in a frequency shift of the plate mechanical resonance frequency $\omega$ given by the expression
\begin{equation}
\tilde{\omega}^2=\omega^2+k_{\textrm{air}}=\omega^2+\frac{P}{\rho t d}.
\label{eq:kair}
\end{equation}
We consider the vibrations of the clamped membranes in the direction orthogonal to their plane ($x$-direction, see Fig.~\ref{fig:sandwich}a) and denote by $x_1$ and $x_2$ the amplitudes of two corresponding normal modes with frequencies $\omega_1$ and $\omega_2$ and intrinsic dampings (in vacuum) $\gamma_1$ and $\gamma_2$, respectively. If only one membrane, say 1, was oscillating while the other was fixed, the equation of motion for the normal mode considered would read
\begin{equation}
\ddot{x}_1+(\gamma_1+\gamma_\textrm{air})\dot{x}_1+\omega_1^2x_1+k_\textrm{air}x_1=F_1,
\end{equation}
with $\gamma_{\textrm{air}}$ given by Eq.~(\ref{eq:gamma_air}) and $F_1$ the noise force associated to the thermal fluctuations and the collisions with air molecules.

In the case of a sandwich consisting of two parallel, nearly identical oscillating membranes, the squeeze film forces are opposite on each membrane, resulting in the coupled dynamical equations~\cite{Naesby2017}
\begin{eqnarray}
\ddot{x}_1+(\gamma_1+\gamma_{\textrm{air}})\dot{x}_1+\omega_1^2x_1+k_{\textrm{air}}(x_1-x_2)=F_1,\label{eq:x1}\\
\ddot{x}_2+(\gamma_2+\gamma_{\textrm{air}})\dot{x}_2+\omega_2^2x_2+k_{\textrm{air}}(x_2-x_1)=F_2,\label{eq:x2}
\end{eqnarray}
with $F_2$ the noise force for mode 2. Note that, in writing Eqs.~(\ref{eq:x1}-\ref{eq:x2}), we assumed that modes 1 and 2 have similar frequencies and neglected the contributions due to  other vibrational modes, whose frequencies were assumed to be very different and whose off-resonant contributions to the frequency shifts of modes could then be neglected, as is discussed below.

\begin{figure}[h]
\centering\includegraphics[width=0.9\columnwidth]{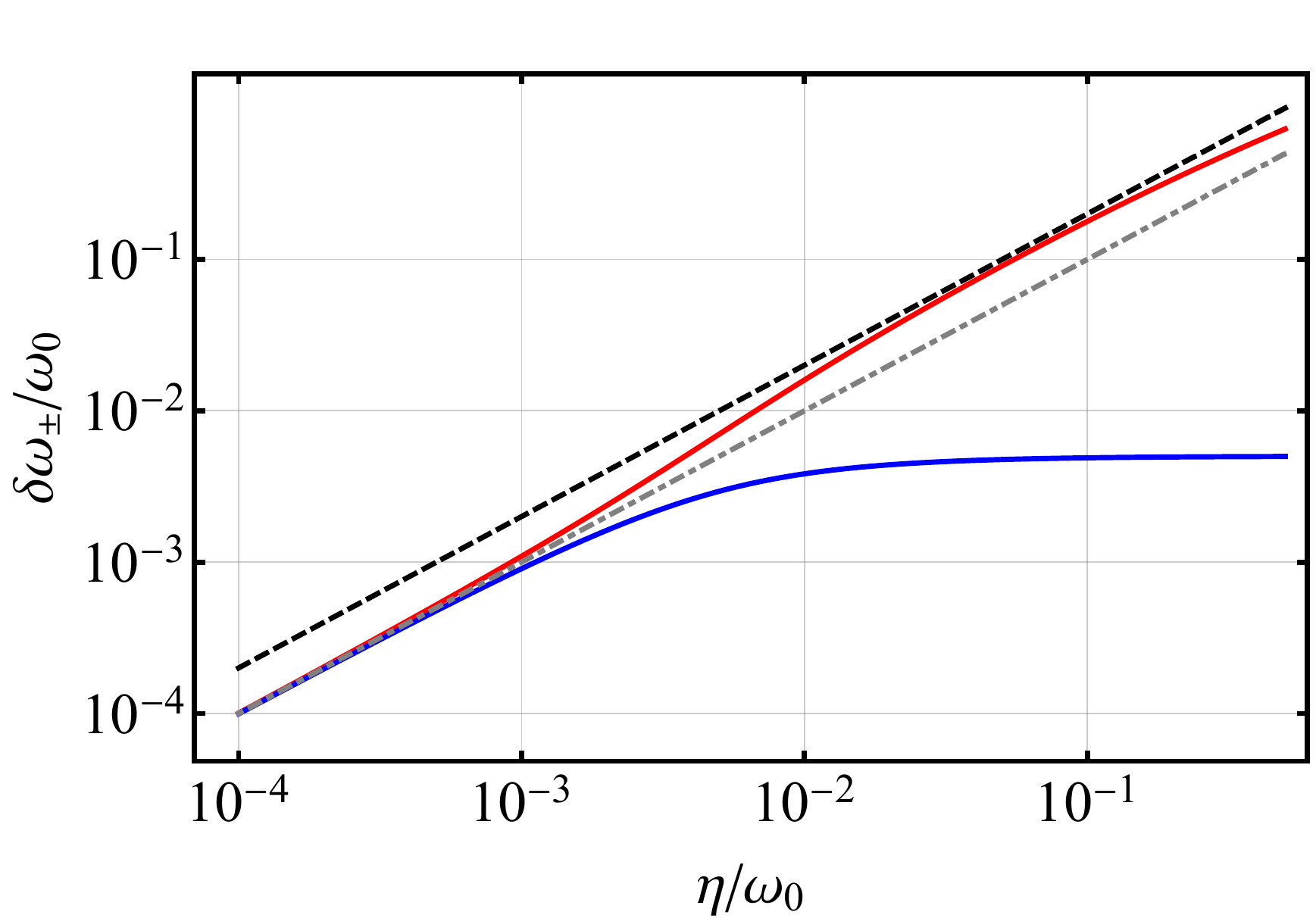}
\caption{Normal mode frequency shifts as a function of air-induced coupling (in units of $\omega_0$) in the case $\delta=0.005\omega_0$. Plain red: $\delta\omega_+$. Plain blue: $\delta\omega_-$. The dot-dashed gray and the dashed black lines show linear shifts equal to $\eta$ and $2\eta$, respectively.}
\label{fig:coupled}
\end{figure}

The dynamics of the pressure-coupled modes 1 and 2, and thereby their thermal noise spectra, depend on the relative strength of the air-induced coupling, their frequency separation and their respective (pressure-dependent) dampings. As observed in~\cite{Naesby2017}, very non-degenerate frequency modes behave independently of each other and their mechanical resonance frequencies are both shifted by a nearly equal amount at a given pressure according to Eq.~(\ref{eq:kair}). For nearly degenerate frequency modes, however, hybridization occurs and normal "bright" and "dark" modes have to be defined~\cite{Naesby2017}, due to the intermode coupling provided by the squeezed gas. More quantitatively, we consider the good oscillator limit---i.e. assume that the mechanical quality factors $Q_i=\omega_i/(\gamma_i+\gamma_{\textrm{air}})$ ($i=1,2$) remain large over the pressure range considered. The frequencies of these normal modes can be found by analyzing the Fourier transforms of Eqs.~(\ref{eq:x1}-\ref{eq:x2}). One obtains:
\begin{equation}
\omega_{\pm}=\left[\omega_0^2+\delta^2+2\eta\omega_0 \pm 2\omega_0\sqrt{\delta^2+\eta^2}\right]^{1/2},
\end{equation}
where 
\begin{equation}\omega_0=\frac{\omega_1+\omega_2}{2},\hspace{0.2cm} \delta=\frac{\omega_1-\omega_2}{2},\hspace{0.2cm}\textrm{and}\hspace{0.2cm} \eta=\frac{k_{\textrm{air}}}{2\omega_0}.\label{eq:omega0deltaeta} \end{equation}
 Figure~\ref{fig:coupled} illustrates the variations of the frequency shifts, $\delta\omega_{\pm}=\omega_{\pm}-(\omega_0\pm\delta)$, as the air-induced coupling $\eta$ (proportional to the pressure) is varied, in the case of a positive $\delta=0.005\omega_0$. Two regimes can be distinguished: when $\eta\ll\delta$, the dynamics of both modes are essentially independent and they experience the same positive frequency shift $\eta$. As the coupling is increased, the bare modes hybridize and, when $\eta\gg\delta$, the bright mode---corresponding to the relative motion of the membranes---experiences a doubled frequency shift, $2\eta$, while the dark mode---corresponding to their center-of-mass motion---no longer experiences any pressure shift and its frequency converges to the mean of the bare frequencies $\omega_0$. At very large couplings the bright mode frequency shift no longer scales linearly with $\eta$, but rather as $\sqrt{\eta}$. When $\delta$ is negative the roles of $\omega_+$ and $\omega_-$ are reversed.

\section{SiN membrane sandwich}
\label{sec:sandwich}

\subsection{Assembly}

\begin{figure}[h]
\centering\includegraphics[width=\columnwidth]{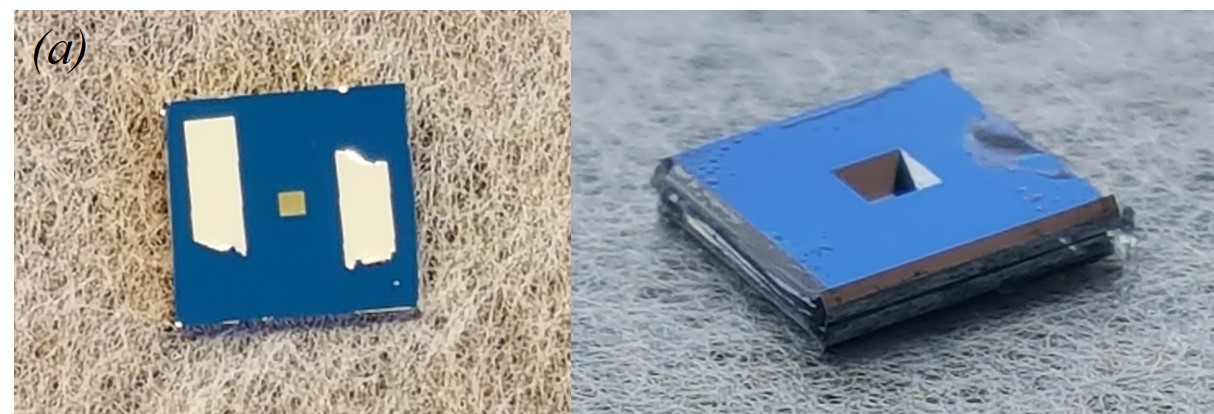}\\\vspace{0.2cm}
\centering\includegraphics[width=0.9\columnwidth]{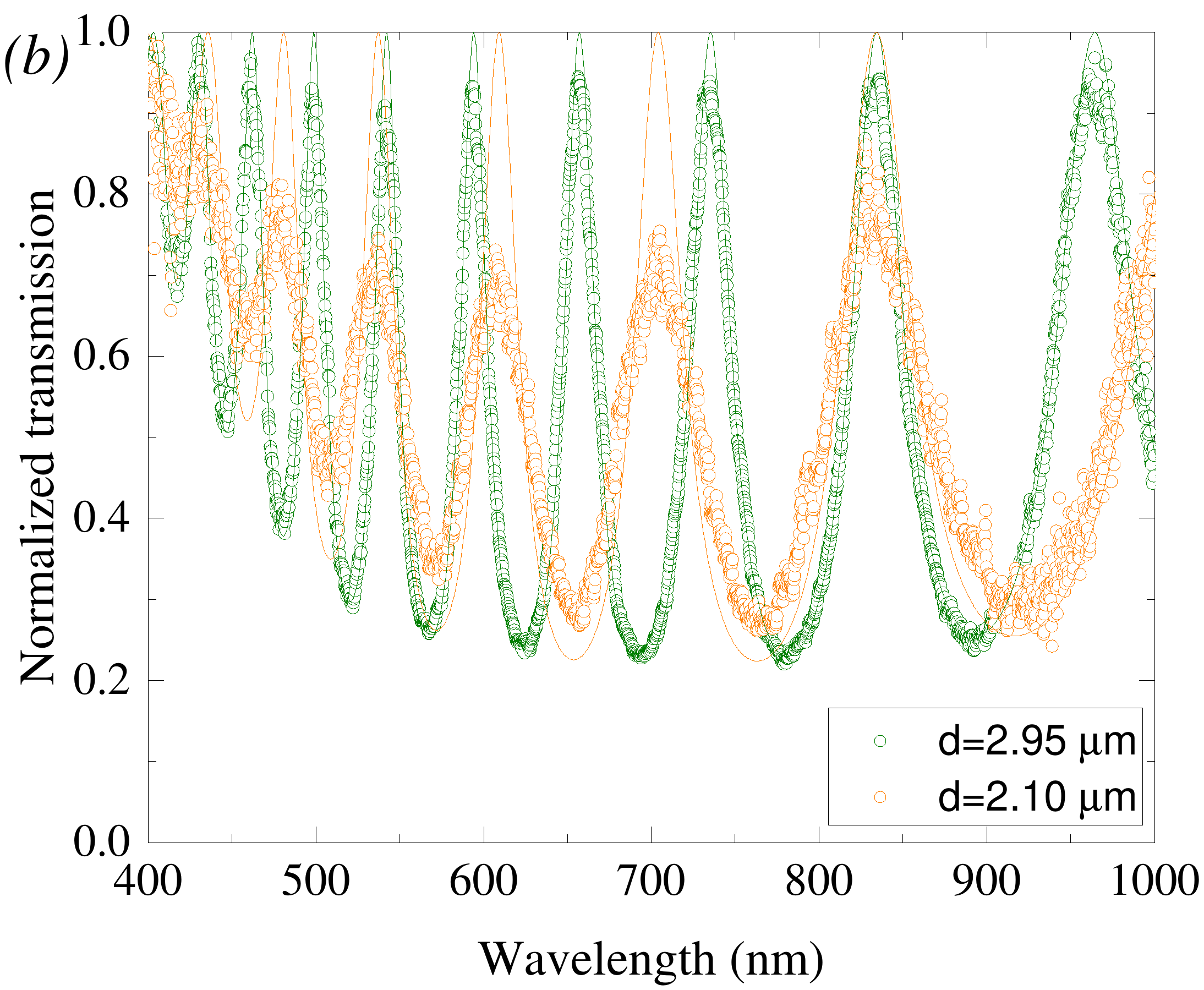}
\caption{(a) Pictures of the membrane chip with Al spacer used in Sec.~\ref{sec:AL} (left) and the assembled sandwich chip (right). The membrane is clamped on all sides to the frame and gas between the membranes can leak through the gap between the frames on the sides where no spacer is present. (b) Normalized transmission spectra of sandwiches with $d=2.95$ $\mu$m (green) and $d=2.10$ $\mu$m (orange) under broadband illumination. The solid lines indicate the results of fits with the theoretical model discussed in the text.}
\label{fig:assembly}
\end{figure}

The SiN membranes used in this work are commercial (Norcada Inc.), high-tensile stress ($\sim $ 0.9 GPa), 500 $\mu$m-square and 87 nm-thick stochiometric films deposited on a 5 mm-square and 500 $\mu$m-thick Si frame. Two approaches were used for the sandwich assembly: in the first one, two rectangular, 1 $\mu$m-thick aluminium spacers ($0.5\times 1$ mm$^2$) were deposited onto one chip about 1 mm from the suspended membrane (Fig.~\ref{fig:assembly}a). The upper membrane was then positioned parallel to the lower one using piezoactuators while being illuminated by a broadband light source and the spectrum of the transmitted light through the array being monitored using a fiber spectrometer~\cite{Nair2017}. Small dabs of UV resist (OrmoComp, Micro resist technology GmbH) were deposited on the sides of the frame and the resist cured when reasonable parallelism was achieved. In the second approach, a thin line of UV resist deposited close to two edges of the lower chip played the role of a flexible spacer. The upper chip was then gently pressed down using the piezoactuators in the same way as previously before the resist was cured. Figure~\ref{fig:assembly} shows a picture of one such sandwich, as well as examples of normalized transmission spectra of two sandwiches with $d=2.10$ and $d=2.95$ $\mu$m under broadband illumination and in air. The solid lines are fits to the transmission function of a linear Fabry-Perot etalon, which takes into account the refractive index of SiN, the membrane thickness $t$ and the intermembrane separation $d$~\cite{Nair2016,Nair2017}. From these fits the intermembrane separation $d$ and thickness $t=87(1)$ nm can be accurately determined. The reduced contrast in the interference fringes of the shorter sandwich is attributed to a rather poor degree of parallelism (estimated tilt angle $\sim$ 2 mrad) achieved after assembly, which has bearings on the squeeze film effect for this sample, as will be discussed later.

\subsection{Optomechanical characterization}

\begin{figure}[h]
\centering\includegraphics[width=0.75\columnwidth]{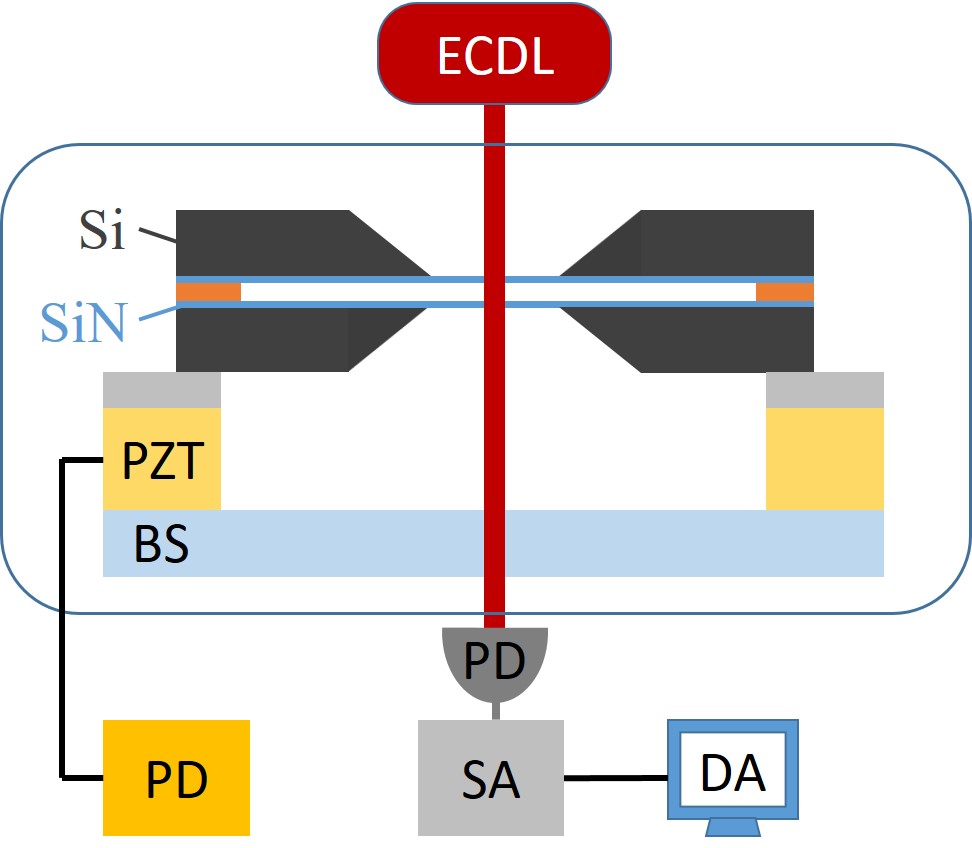}
\caption{Schematic of the optomechanical characterization setup. ECDL: external cavity diode laser, PZT: piezoelectric transducer, BS: beamsplitter, PD: photodiode, SA: spectrum analyzer, PD: PZT driver, DA: data analysis computer.}
\label{fig:setup}
\end{figure}

In order to characterize the mechanical properties of the assembled sandwiches, the samples are lying in a 450 cm$^3$ vacuum chamber on a ring-shaped mount, the corners of the lower chip resting on the ring. A 50:50 beamsplitter, placed approximately 7 mm away from the sample, forms a linear Fabry-Perot interferometer whose length is adjustable with a piezoelectric transducer (Fig.~\ref{fig:setup}). The transmission of monochromatic light issued from a tunable external cavity laser diode ($\sim 900$ nm) is recorded with a fast photodiode. The length of the interferometer is adjusted so as to  maximize the displacement sensitivity (typically close to the interference signal midfringe) and the fluctuations of the photocurrent are analyzed using a low resolution bandwidth spectrum analyzer. The thermal noise spectrum--i.e. the noise spectrum in absence of external driven force--is typically recorded over a span range of a few to a few hundreds of kilohertz with a resolution bandwidth of 0.5 Hz and averaged 500 times. The vacuum chamber temperature not being actively stabilized, resonance frequency drifts of  few tens of hertz per hour are typically observed and corrected for during long acquisition time measurement series.
The pressure is controlled by opening the valve to the chamber and measured using an ion gauge sensor, calibrated against a Pirani gauge whose air pressure responsitivity was absolutely calibrated in the range 1-50 mbar~\cite{Jakobsen2019}. The calibrated Pirani sensor response is accurate to within a few percent in that range and its response is used to calibrate that of the measuring ion gauge sensor.

\section{Experimental results}
\label{sec:results}

\subsection{2.95 $\mu$m, Al-spacer sandwich}\label{sec:AL}

\begin{figure}[h]
\centering\includegraphics[width=0.9\columnwidth]{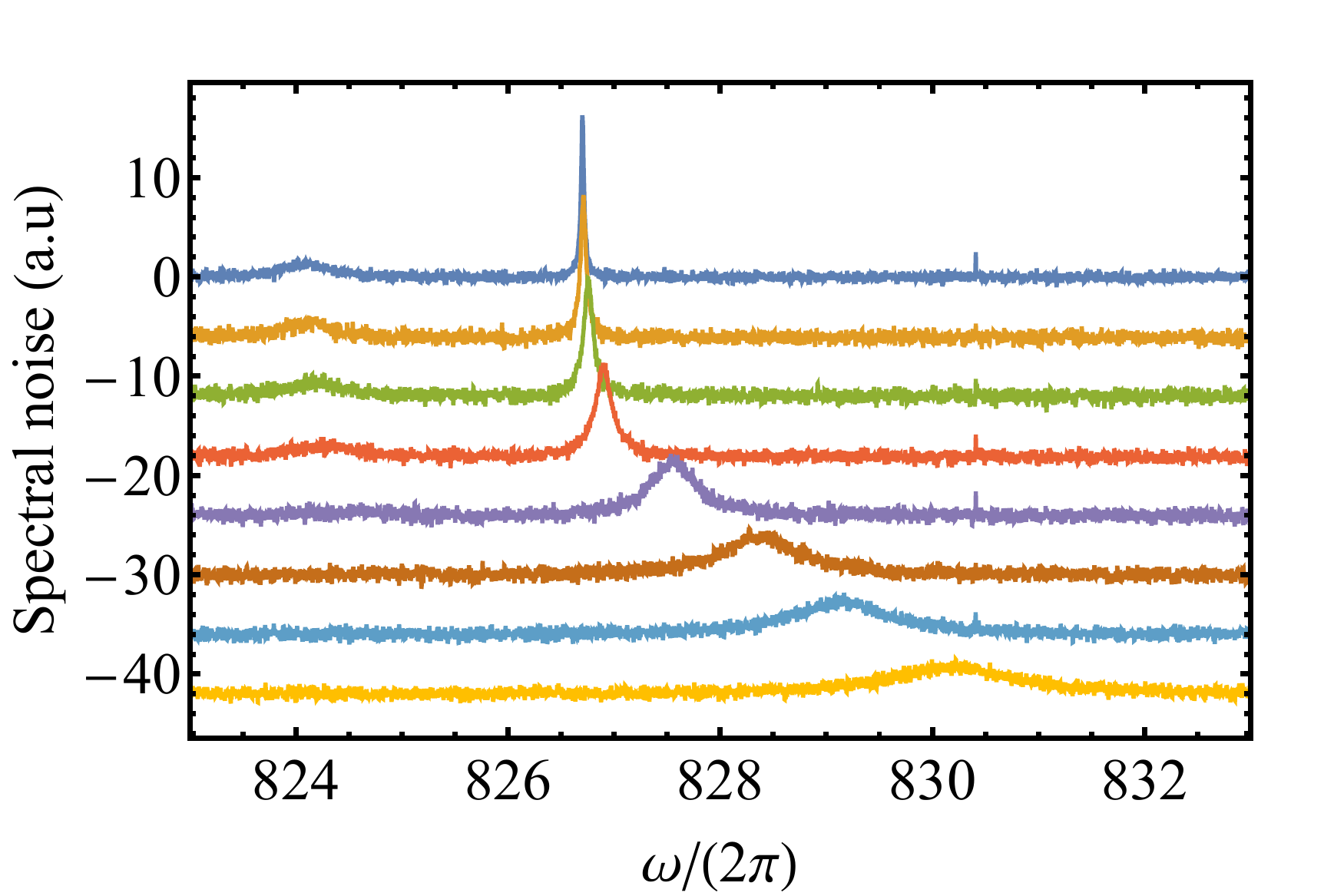}
\caption{Thermal noise spectra around the fundamental mode frequencies for different pressures (from top to bottom: $P=5\;10^{-8}$, 0.018, 0.055, 0.13, 0.35, 0.6, 0.82, 1.05 mbar). The background-subtracted spectra are shown with a logarithmic scale and are vertically offset for clarity.}
\label{fig:AL_spectra}
\end{figure}

\begin{figure}[h]
\centering\includegraphics[width=0.9\columnwidth]{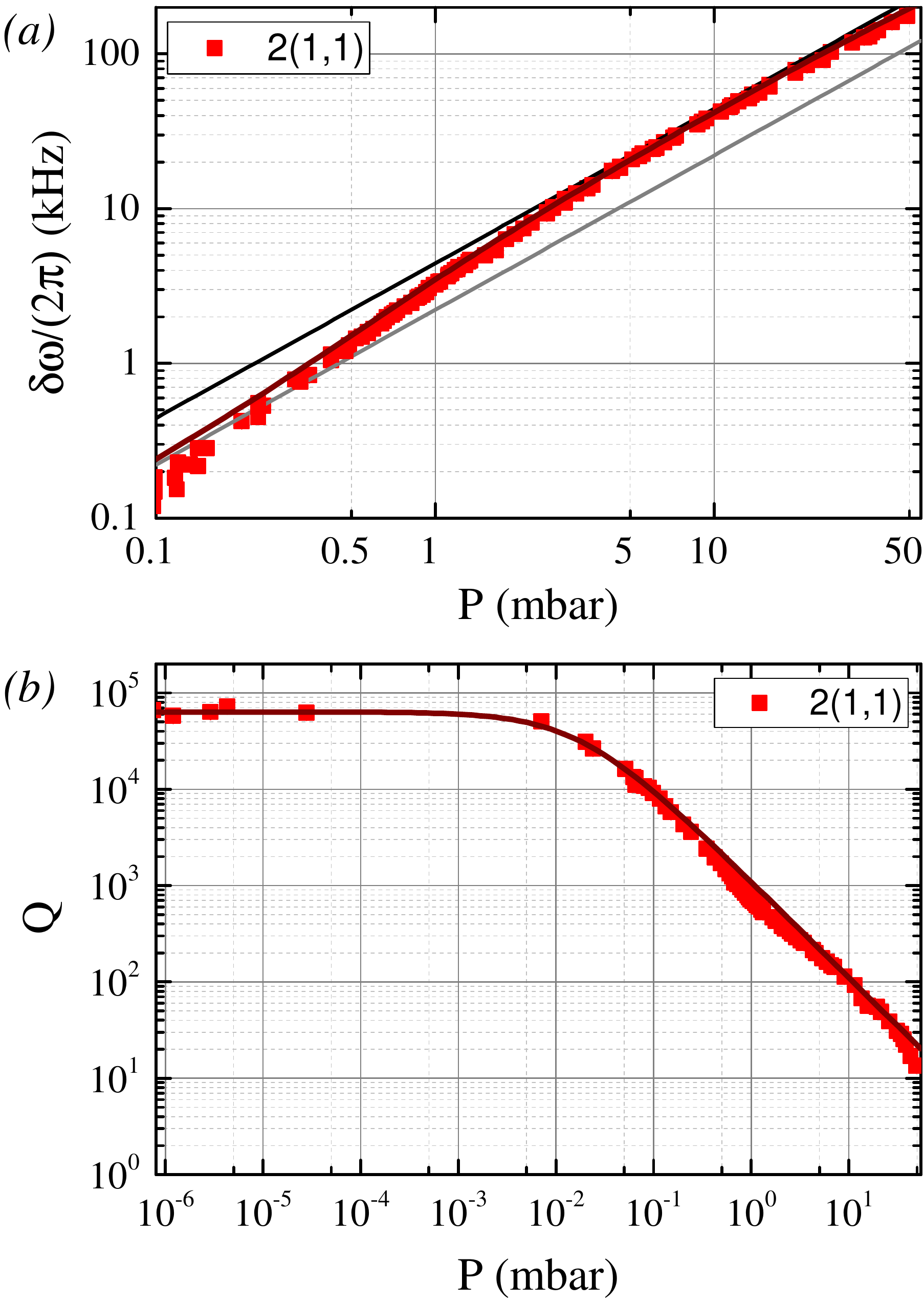}
\caption{Resonance frequency shift (a) and mechanical quality factor (b) of membrane 2's fundamental mode as a function of pressure (logarithmic scale).  The solid dark red line in (a) shows the predictions of the squeeze film model. The single- and double-linear shifts discussed and shown in Fig.~\ref{fig:coupled} are shown by the gray and black lines, respectively. The solid dark red line in (b) shows the kinetic damping predictions.}
\label{fig:AL}
\end{figure}

We first characterize the pressure response of a sandwich with aluminium spacers and $d=2.95$ $\mu$m. While one membrane (membrane 2) was found to exhibit good quality factor modes, the other membrane showed modes with substantially reduced Qs, which could be the result of the Al deposition or manipulation in the course of the deposition process. The fundamental mode frequencies of both membranes were $\omega_1/(2\pi)=824.1$ kHz and $\omega_2/(2\pi)=826.7$ kHz, and their intrinsic quality factors $Q_1=1500$ and $Q_2=65000$, respectively. Figure~\ref{fig:AL_spectra} shows examples of thermal noise spectra around these frequencies for different pressures. Positive frequency shifts and broadening of the spectra with increasing pressure are clearly observed for both modes. Due to its lower Q the thermal fluctuations of membrane 1's fundamental mode can no longer be resolved for pressures above 0.3 mbar, however. In contrast, membrane 2's thermal fluctuations can be resolved up to much higher pressures and shifts as high as 176 kHz were measured at 50 mbars (see Fig.~\ref{fig:AL}a).

Figure~\ref{fig:AL} shows the variations with pressure of the mechanical quality factor $Q=\omega/\gamma$ (where $\gamma/(2\pi)$ is the full width at half maximum of the noise spectrum) and the resonance frequency shift $\delta\omega/(2\pi)$ 
of membrane 2's fundamental mode, denoted by 2(1,1) (where 2 stands for membrane 2 and (1,1) for $m=n=1$). The quality factors and shifts were extracted from Lorentzian fits to thermal noise spectra as shown above. The observed frequency shifts are observed to be in good agreement with the predictions of the coupled-oscillator squeeze film model given the independently measured intermembrane distance $d=2.95$ $\mu$m. The crossover from independent to coupled modes predicted by the squeeze film model discussed in Sec.~\ref{sec:squeeze} is also clearly visible, resulting in a pressure responsivity of about 4 kHz/mbar in the range 2-20 mbar. The measured quality factors are also observed to be in good agreement with the kinetic damping predictions, confirming that extra damping due to the essentially elastic squeeze film force is negligible with respect to kinetic damping. 

\begin{figure}[h]
\centering\includegraphics[width=0.9\columnwidth]{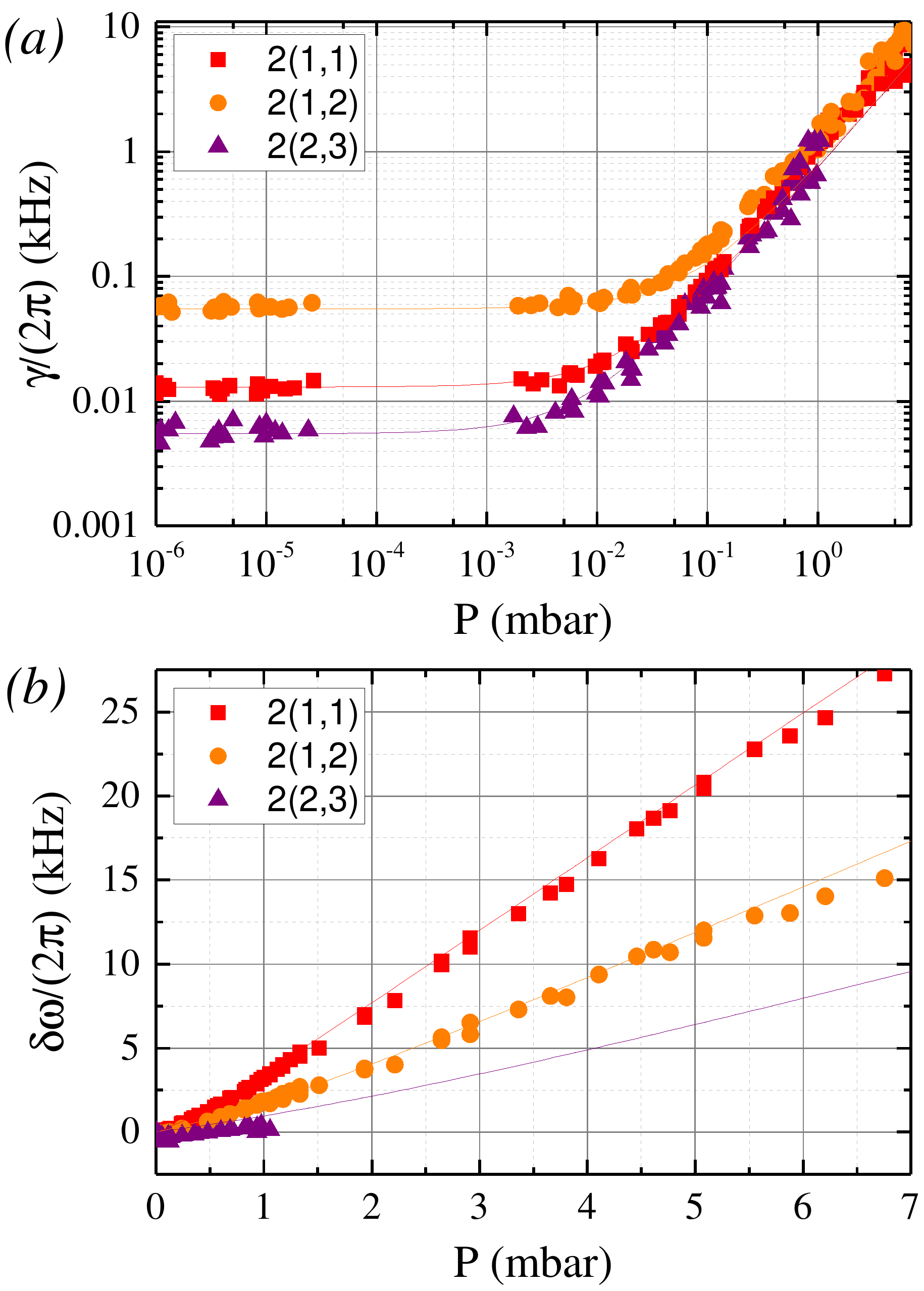}
\caption{Damping rates (a) and resonance frequency shifts (b) versus pressure for three modes of membrane 2. Red squares: (1,1) mode, orange circles: (1,2) mode, purple triangles: (2,3) mode. The solid lines show the theoretical predictions.}
\label{fig:ALmodes}
\end{figure}

To further ascertain this, the effects of pressure on the damping rate and resonance frequency of two higher-order modes, the (1,2) mode with frequency of 1306 kHz and intrinsic Q-factor of 24000 and the (2,3) mode with frequency of 2108 kHz and intrinsic Q-factor of 383000, were also investigated. The results, shown in Fig.~\ref{fig:ALmodes}, confirm that the variations of the damping rates with pressure above 10$^{-3}$ mbar are frequency-independent, as expected from kinetic damping, while the frequency shift pressure responsivity is inversely proportional with the resonance frequency, as expected from the squeeze film model predictions ($\eta\varpropto 1/\omega_0$, see Eq.~(\ref{eq:omega0deltaeta})).

\subsection{2.1 $\mu$m, UV resist-spacer sandwich}

\begin{figure}[h]
\centering\includegraphics[width=0.9\columnwidth]{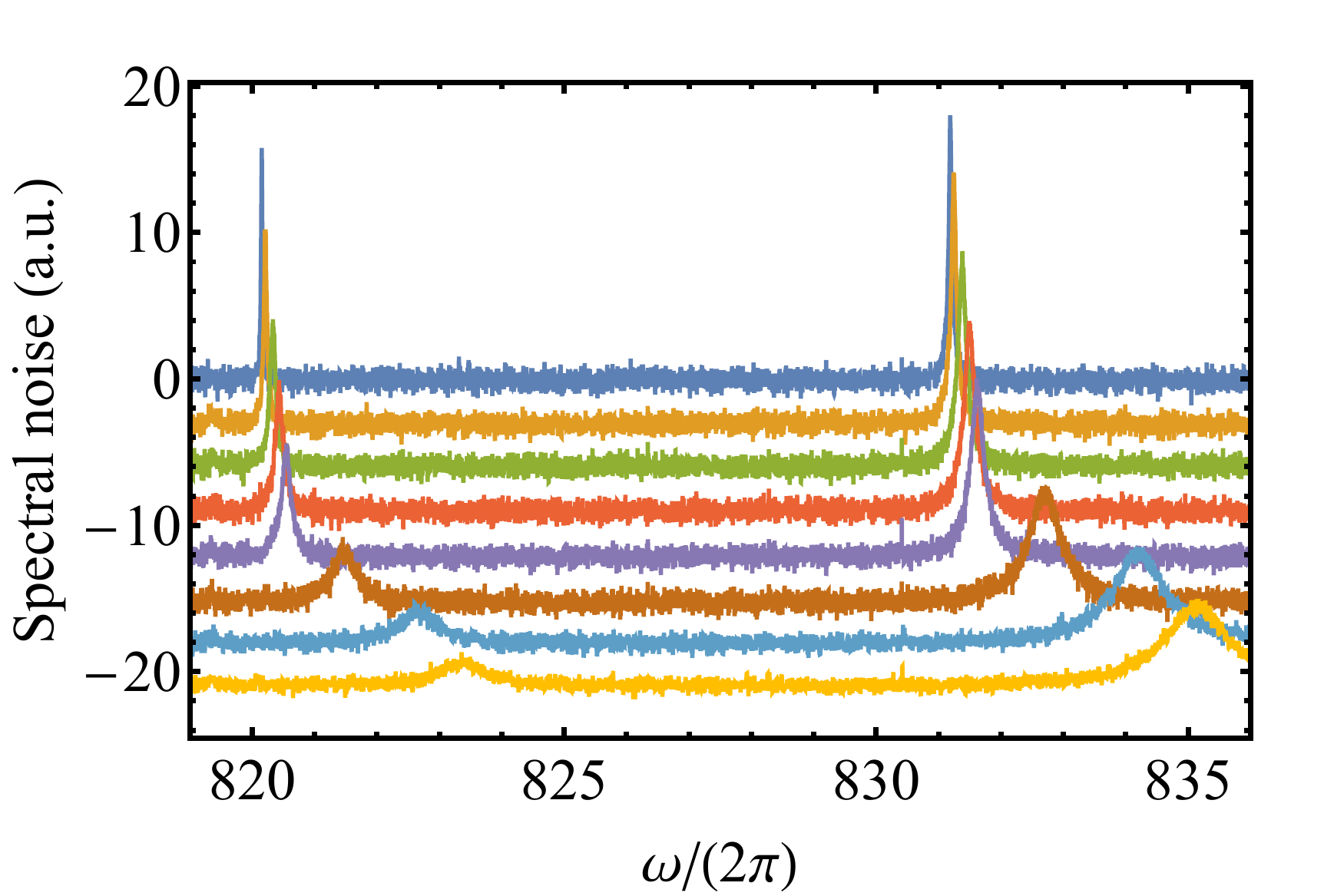}
\caption{Thermal noise spectra around the fundamental mode frequencies for different pressures (from top to bottom: $P=1.3\;10^{-4}$, 0.02, 0.07, 0.11, 0.15, 0.47, 0.89, 1.12  mbar). The background-subtracted spectra are shown with a logarithmic scale and vertically offset for clarity.}
\label{fig:S10_spectra}
\end{figure}

\begin{figure}[h]
\centering\includegraphics[width=0.9\columnwidth]{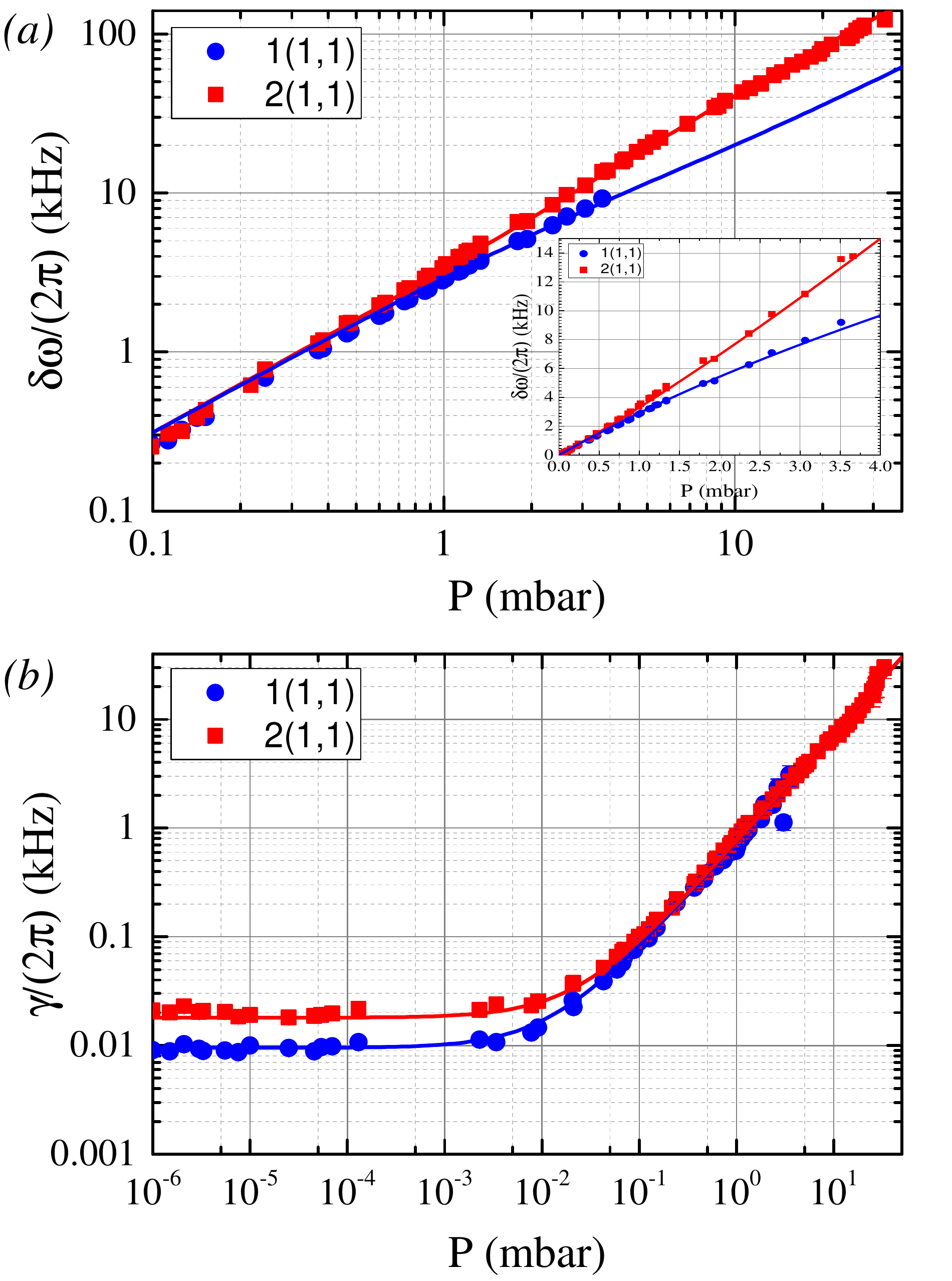}
\caption{Frequency shifts (a) and damping (b) of both membranes' fundamental modes as a function of pressure (logarithmic scale). The solid lines in (a) indicate the predictions of the two-spring squeeze model discussed in the text. The solid lines in (b) indicate the kinetic damping predictions. The inset in (a) shows a zoom-in (linear scale) of the frequency shifts versus pressure in the few millibar range.}
\label{fig:S10}
\end{figure}

We now turn to a second membrane sandwich, this time assembled using the UV resist-spacer approach. For this sample, whose intermembrane separation is slightly smaller ($d=2.1$ $\mu$m) than the previous one, higher quality factor modes were observed for both membranes, which may be an indication that this assembly method does not increase clamping losses as much as the first one. The fundamental modes of both membranes have frequencies $\omega_1/(2\pi)=820.1$ kHz and $\omega_2/(2\pi)=831.2$ kHz, with intrinsic Q factors of 86000 and 46000, respectively. Examples of thermal noise spectra at different pressures are shown in Fig.~\ref{fig:S10_spectra} and the variations of the resonance frequency shifts and damping rates of both modes with pressure are shown in Fig.~\ref{fig:S10}. Kinetic damping broadening of the spectra and large positive resonance frequency shifts are also observed with this sample. At low pressures, both modes experience similar linear frequency shifts with a responsivity of 3 kHz/mbar, as expected from the squeeze film predictions. However, the resonance frequency shifts in the crossing region do not vary in exactly the same way as with the previous sample. moreover, the lower frequency mode still exhibits a noticeable frequency shift at high pressure, while the high frequency mode sees its responsivity increase, but less than by a factor of 2. 

This qualitatively different behavior can be accounted for by considering that, while a fraction of the squeeze film-added spring constant couples the two modes together as previously, the remaining fraction affects each mode independently. The latter interaction, corresponding to each membrane being independently spring-coupled to the other one behaving as a "fixed" plate, can be reasonably envisaged if the modes of each membranes are transversely offset with respect to each other or if the membranes are not parallel, as we strongly surmize is the case for this sandwich. Assuming for simplicity the same membrane-"fixed" plate spring constant for both modes, Eqs.~(\ref{eq:x1})-(\ref{eq:x2}) are modified to:
\begin{eqnarray}
\ddot{x}_1+(\gamma_1+\gamma_{\textrm{air}})\dot{x}_1+\omega_1^2x_1+k'_{\textrm{air}}(x_1-x_2)+k''_{\textrm{air}}x_1=F_1,\label{eq:x1bis}\\
\ddot{x}_2+(\gamma_2+\gamma_{\textrm{air}})\dot{x}_2+\omega_2^2x_2+k'_{\textrm{air}}(x_2-x_1)+k''_{\textrm{air}}x_2=F_2,\label{eq:x2bis}
\end{eqnarray}
where $k'_{\textrm{air}}$ and $k''_{\textrm{air}}$ are the air-induced membrane-membrane and membrane-"fixed" plate spring constants, respectively. The normal mode frequency shifts are accordingly given \red{by:}
\begin{equation}
\omega_{\pm}=\left[\omega_0^2+\delta^2+2(\eta'+\eta'')\omega_0 \pm 2\omega_0\sqrt{\delta^2+\eta'^2}\right]^{1/2},
\label{eq:omegapmbis}
\end{equation}
where $\eta'=k'_{\textrm{air}}/(2\omega_0)$ and $\eta''=k''_{\textrm{air}}/(2\omega_0)$ are both proportional to the pressure, as previously. At low pressures (as long as $\eta',\eta''\ll\delta$) both modes experience the same linear frequency shift $\eta'+\eta''$. At high pressures (when $\eta',\eta'\gg\delta'$) the higher frequency mode experiences a shift given by $2\eta'+\eta''$, while the lower frequency mode experiences a shift $\eta''$. The solid lines in Fig.~\ref{fig:S10}a show the predictions of Eq.~(\ref{eq:omegapmbis}) when $\eta'+\eta''$ is equal to the full shift $\eta$ given by the squeeze film model [Eq.~(\ref{eq:kair})] for $d=2.1$ $\mu$m and when $\eta'\simeq\eta''$; these predictions match well the experimental data in both the crossover and high pressure regions. The slight apparent discrepancy at pressures around 0.1 mbar is most likely the result of thermal drifts during the measurement series, issue which we now turn to.

\begin{figure}[h]
\centering\includegraphics[width=0.9\columnwidth]{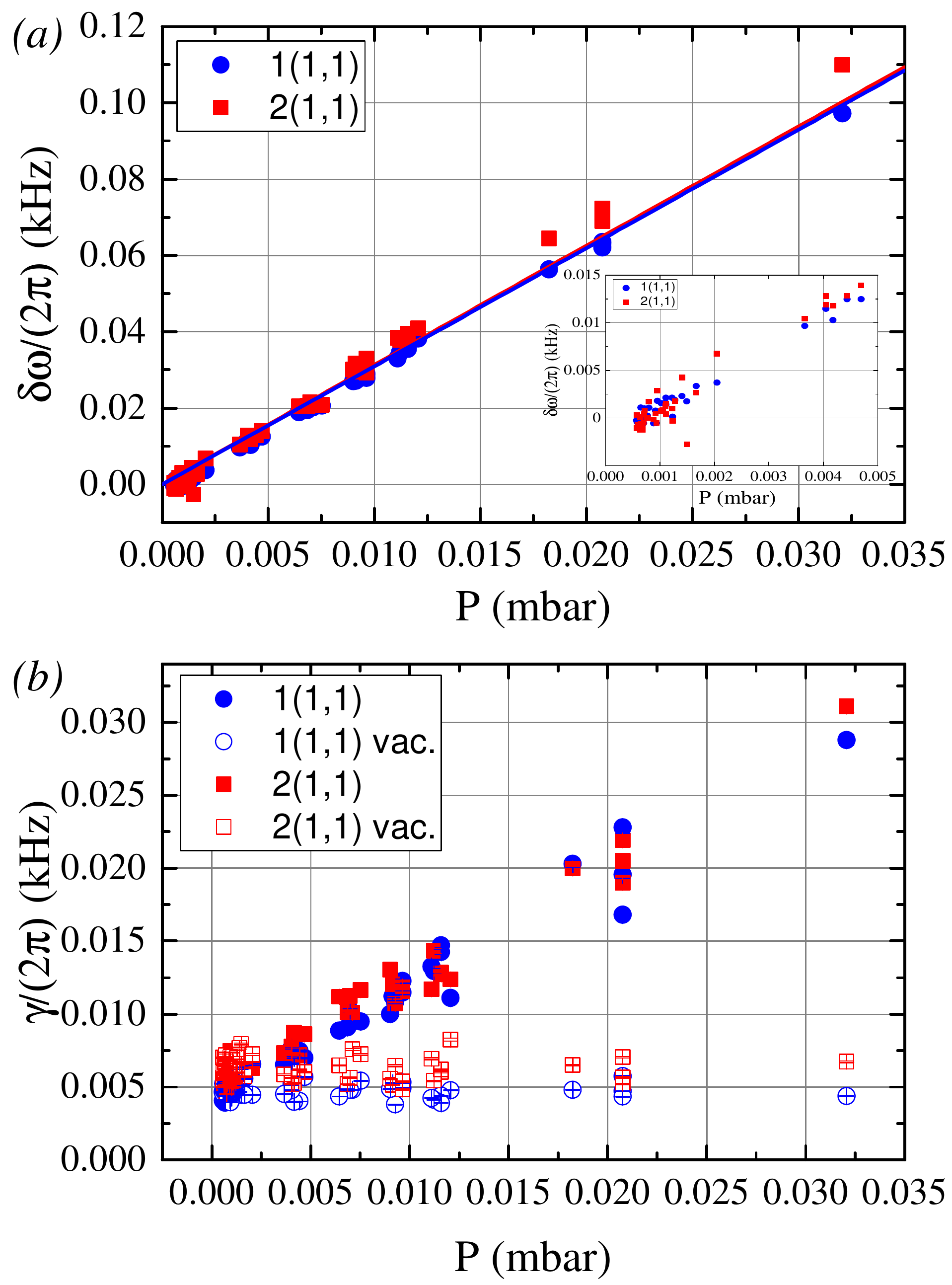}
\caption{(a) Frequency shifts of both membranes' fundamental modes as a function of pressure. The inset shows a zoom into the few $10^{-3}$ mbar region. (b) Corresponding linewidths (full symbols). The empty symbols show the linewidths measured in the corresponding sequence in high vacuum (below $10^{-4}$ mbar). For all data the Lorentzian fit result uncertainties are smaller than the size of the symbols.}
\label{fig:S10low}
\end{figure}

In order to estimate the pressure sensitivity of our sensors in the sub-millibar range and to be less sensitive to the thermal drifts in the chamber, we carried out sequential measurements during which the air pressure was first quickly increased from high vacuum ($\sim 10^{-5}$ mbar) to a certain value in the sub-millibar range, the thermal noise spectrum acquired, and the pressure quickly decreased again to below $10^{-4}$ mbar and the thermal noise spectrum measured again. Repeating this sequence for various pressures in the range $10^{-3}-10^{-1}$ mbar yield the frequency shifts shown in Fig.~\ref{fig:S10low}a. Equal frequency shifts with a responsivity of 3.1 kHz/mbar are observed for each mode and in good agreement with the theoretical expectations. Zooming into the few $10^{-3}$ mbar region (inset of Fig.~\ref{fig:S10low}a) shows that the current pressure sensitivity using the squeeze film frequency shift is at the 0.1 Pa level. Let us note that it should be possible to improve this sensitivity substantially by active temperature stabilization of the chamber, use of lower frequency/higher Q modes and further reduction of the intermembrane distance.

In principle, the kinetic damping broadening of such high Q mechanical resonances can also be used to infer pressure as per Eq.~(\ref{eq:gamma_air}), although in a species dependent way. Figure~\ref{fig:S10low}b shows the variations in the linewidth of the thermal noise spectrum of each mode at various pressures in the same range, as well as the corresponding linewidths measured in high vacuum during the same sequence. The linewidth measurements typically reveal an overall larger statistical spread than the frequency shift measurements. Given the lower pressure responsivity of the broadening ($\gamma_{\textrm{air}}/P\simeq (2\pi)0.7$ kHz/mbar) than that provided by the resonance frequency shift, the resulting sensitivity is consequently found to be slightly worse. The pressure sensitivity to air damping could in principle be increased, though, by a reduction of the statistical spread in the linewidth determination (using e.g. mechanical ringdown spectroscopy techniques) and the use of higher Q mechanical resonances.

\section{Conclusion}
\label{sec:conclusion}

The effects of air pressure on SiN membrane sandwiches with gaps in the 2-3 micron range were investigated in the rarefied air and transition regimes  ($10^{-3}-50$ mbar), via the measurement of their thermal noise spectra by optical interferometry. The essentially elastic squeeze film force due to the compression of the gas between the membranes results in strong positive shifts of the mechanical resonance frequencies of the membranes, which can be enhanced by the air-induced coupling between the membrane modes. The experimental observations are in good agreement with a simple coupled-oscillator model which includes both squeeze film and kinetic damping effects. The high pressure responsivity (several kHz/mbar) and (sub-pascal) sensitivity exhibited by these squeeze film sensors are substantially improved-- by a factor $\sim 3$--over those observed with larger gap sandwiches. The performances of these sensors could be further enhanced by e.g. reducing the intermembrane distance, using higher Q mechanical resonances and better temperature stability or interferometric displacement sensitivity, which would make them attractive for direct and absolute pressure measurements in rarefied air and high vacuum environments.

\begin{acknowledgements}
We are grateful to Bjarke R. Jeppesen for his help with the aluminium spacer deposition and to Jonathan Merrison for his help with the calibrated pressure gauge. We also acknowlegde support from Villum Fonden.
\end{acknowledgements}

\bibliography{pressure_sensor_bib}

\end{document}